\documentclass[12pt]{article}
\usepackage{graphicx}
\usepackage[cp1251]{inputenc}

\begin{document}
 \noindent {\footnotesize\it Astronomy Letters, 2019, Vol. 45, No 1, pp. 10--19.}
 \newcommand{\dif}{\textrm{d}}

 \noindent
 \begin{tabular}{llllllllllllllllllllllllllllllllllllllllllllll}
 & & & & & & & & & & & & & & & & & & & & & & & & & & & & & & & & & & & & & \\\hline\hline
 \end{tabular}

  \vskip 0.5cm
  \centerline{\bf\large Parameters of the Link between the Optical and Radio Frames}
  \centerline{\bf\large from Gaia DR2 Data and VLBI Measurements}
  \bigskip
  \bigskip
  \centerline
 {V.V. Bobylev\footnote [1]{e-mail: vbobylev@gaoran.ru} }
  \bigskip

  \centerline{\small\it Pulkovo Astronomical Observatory, Russian Academy of Sciences,}

  \centerline{\small\it Pulkovskoe sh. 65, St. Petersburg, 196140 Russia}
 \bigskip
 \bigskip
 \bigskip

 {
{\bf Abstract}---Based on published data, we have assembled a
sample of 88 radio stars for which there are both trigonometric
parallax and proper motion measurements in the Gaia DR2 catalogue
and VLBI measurements. A new estimate of the systematic offset
between the optical and radio frames has been obtained by
analyzing the GaiaDR2--VLBI trigonometric parallax differences:
 $\Delta\pi=-0.038\pm0.046$~mas (with a dispersion of 0.156 mas). This
means that the Gaia DR2 parallaxes should be increased by this
correction. The parallax scale factor is shown to be always very
close to unity within $\sim$3 kpc of the Sun: $b=1.002\pm0.007.$
Our analysis of the proper motion differences for the radio stars
based on the model of solid-body mutual rotation has revealed no
rotation components differing significantly from zero:
 $(\omega_x,\omega_y,\omega_z)=(-0.14,0.03,-0.33)\pm(0.15,0.22,0.16)$ mas yr$^{-1}.$
  }

 %\medskip DOI: 10.1134/S1063773718110026

 \subsection*{INTRODUCTION}
Highly accurate stellar parallaxes are required to solve many
stellar-astronomy problems. The trigonometric parallaxes are among
the most reliable ones. However, it is necessary to check and
eliminate the possible systematic offsets before using even the
most reliable data.

The first data release of the Gaia space experiment was published
in September 2016 (Prusti et al. 2016; Brown et al. 2016). The
second data release of this experiment, Gaia DR2, appeared in
April 2018 (Brown et al. 2018). This catalogue contains the
trigonometric parallaxes and proper motions of $\sim$1.7 billion
stars. The derivation of their values is based on the orbital
observations performed over 22 months. The mean errors of the
trigonometric parallax and both stellar proper motion components
in this catalogue depend on the magnitude. For example, for bright
stars $(G<15^m)$ the parallax errors lie within the range
0.04--0.02 milliarcseconds (mas), while for faint stars $(G=20^m)$
they are about 0.7~mas.

Lindegren et al. (2018) pointed out the presence of a possible
systematic offset $\Delta\pi=-0.029$~mas in the Gaia DR2
parallaxes with respect to the inertial reference frame. At
present, there are several reliable distance scales a comparison
with which allows, in the opinion of their authors, the
systematics of the Gaia trigonometric parallaxes to be checked.

Stassun and Torres (2016) found quite a significant mean offset
$\Delta\pi=-0.25\pm0.05$~mas of the Gaia DR1 trigonometric
parallaxes with respect to the parallaxes of a calibration sample
of eclipsing binaries. This result was soon confirmed by other
authors based on an analysis of classical Cepheids close to the
Sun (Casertano et al. 2017), the ground-based parallaxes of the
nearest M dwarfs (Jao et al. 2016), and asteroseismology (Huber et
al. 2017).

Having compared the parallaxes of 89 stars from the Gaia DR2
catalogue and calibration eclipsing binary stars, Stassun and
Torres (2018) found a slight offset between the frames
$\Delta\pi=-0.082\pm0.033$~mas. This value is also confirmed by
other authors, in particular, when analyzing Cepheids (Riess et
al. 2018) and asteroseismology (Zinn et al. 2018).

Therefore, the distances to radio stars determined by very long
baseline interferometry (VLBI) are of interest. Here we have in
mind the absolute parallaxes that are absolutized during the
observations using quasars. At present, the VLBI observations
aimed at determining highly accurate trigonometric parallaxes and
proper motions of radio sources, in particular, galactic masers,
are being performed by several research teams.

The accuracy of astrometric VLBI measurements depends on many
factors. For example, estimates of the contributions from the
position of a calibration source, the Earth's orientation, the
antenna position, and the tropospheric delay for radio sources
located at different declinations can be found in Pradel et al.
(2006). Furthermore, the mean VLBI parallax error depends on the
observation frequency: the higher the frequency, the smaller this
error. As a result, the mean VLBI parallax error in observations
at 22.2 GHz is $\sim$0.01 mas.

The Gaia DR2 catalogue contains the astrometric parameters for
more than half a million quasars. This has allowed the optical,
kinematically nonrotating Gaia DR2-Celestial Reference Frame
(Gaia-CRF2) to be realized. Some of the quasars have accurate VLBI
positions, which allows (Mignard et al. 2018) the axes of this
frame to be aligned with the International Celestial Reference
Frame (ICRF) specified by a set of radio sources, for example,
ICRF2 or ICRF3 (which is being developed at present). As Mignard
et al. (2018) showed, the coordinate axes of the Gaia DR2
catalogue and the ICRF3 prototype are aligned with errors of
20--30 mas, but more accurate values of these errors will be
presented after a more detailed study of various errors.
Therefore, determining the mutual rotation parameters between the
two (optical and radio) frames using the proper motions of radio
stars is of interest.

The goal of this paper is to produce a collection of VLBI
observations of the absolute parallaxes and proper motions for
radio stars common to the Gaia DR2 catalogue based on published
data and to use this sample as a calibration one to check the Gaia
DR2 distance scale.

%%%%%%%%%%%%%%%%%%%%%%%%%%%%%%%%%%%%%%%%%%% t1
 \begin{table}[p]
 \caption[]
  {\small Gaia DR2--VLBI stellar proper motion and parallax differences}
  \begin{center}  \label{t:01}  \small
  \begin{tabular}{|l|c|r|r|r|r|r|r|c|c|} \hline
 Star           &  Type or & $\Delta \mu_\alpha\cos\delta,$ & $\sigma_{\Delta\mu_\alpha\cos\delta},$ & $\Delta \mu_\delta,$ & $\sigma_{\Delta\mu_\delta},$ & $\Delta \pi,$ & $\sigma_{\Delta\pi},$ & Ref \\
                & spectrum &      mas yr$^{-1}$   &              mas yr$^{-1}$  &         mas yr$^{-1}$   &               mas yr$^{-1}$   &           mas &                   mas & \\\hline

 SY Scl         &   Mira & $   .541$ &  .328 & $  -.155$ &  .314 & $  -.075$ &  .229 & (1)\\
 S Per          &    RSG & $   .480$ &  .458 & $ -1.380$ &  .451 & $  -.191$ &  .123 & (2)\\
 HII 174        & RS CVn & $   .020$ &  .122 & $  -.105$ &  .172 & $  -.111$ &  .057 & (3)\\ %Плеяды
 HII 625        & BY Dra & $   .409$ &  .134 & $  -.121$ &  .275 & $  -.008$ &  .070 & (3)\\
 HII 1136       & RS CVn & $  -.800$ &  .098 & $   .346$ &  .246 & $  -.161$ &  .057 & (3)\\
 HII 2147       & RS CVn & $ -2.579$ &  .112 & $   .879$ &  .171 & $  -.119$ &  .062 & (3)\\
 V773 Tau       &  T Tau & $ -1.321$ &  .929 & $ -3.935$ &  .391 & $   .113$ &  .164 & (4)\\
 HIP 20097      &  T Tau & $  -.020$ &  .129 & $  -.115$ &  .064 & $  -.084$ &  .059 & (5)\\
 HDE 283572     &  T Tau & $   .158$ &  .150 & $   .106$ &  .134 & $  -.107$ &  .065 & (5)\\
 T Tau N        &  T Tau & $  -.994$ &  .128 & $ -2.037$ &  .112 & $   .109$ &  .066 & (6)\\
 V1201 Tau      &  T Tau & $  -.370$ &  .115 & $ -1.309$ &  .096 & $  -.197$ &  .083 & (5)\\
 V807 Tau       &  T Tau & $   .986$ & 1.237 & $  8.544$ & 1.009 & $   .935$ &  .667 & (5)\\
 V1110 Tau      & RS CVn & $   .438$ &  .112 & $  -.021$ &  .096 & $  -.281$ &  .154 & (5)\\
 HIP 26233      &  B2/3V & $ -1.421$ &  .404 & $  1.311$ &  .409 & $ -2.196$ &  .187 &  (8)\\
 LSI +61 303    &     BH & $  -.146$ &  .041 & $   .185$ &  .067 &       --- &   --- & (7)\\
    DG Tau      &  T Tau & $  -.644$ &  .876 & $  -.197$ &  .932 &       --- &   --- & (9)\\
 HD 118216      &  F2+K2 & $  -.013$ &  .202 & $  -.031$ &  .164 &       --- &   --- & (10)\\
 WR 112         &     WR & $   .625$ & 1.164 & $  1.596$ & 1.436 &       --- &   --- & (11)\\
 WR 125         &     WR & $  -.964$ &  .503 & $   .606$ &  .604 &       --- &   --- & (11)\\
 WR 140         &     WR & $   .377$ &  .206 & $  -.847$ &  .115 &       --- &   --- & (11)\\
 WR 146         &     WR & $  2.284$ &  .696 & $ -1.375$ & 2.242 &       --- &   --- & (11)\\
 WR 147         &     WR & $ -1.097$ &  .803 & $ -1.100$ & 1.198 &       --- &   --- & (11)\\
 PSR J0437--47  & Pulsar & $  1.185$ & 1.198 & $   .654$ & 1.672 & $  1.929$ &  .679 & (12)\\
 V999 Tau       &  T Tau & $ -4.040$ &  .677 & $ -3.306$ &  .433 & $  1.166$ &  .438 & (5)\\
 HD 282630      &  T Tau & $   .410$ &  .191 & $   .078$ &  .152 & $  -.798$ &  .148 & (5)\\
 T Lep          &   Mira & $ -4.887$ &  .555 & $  1.108$ &  .739 & $  -.101$ &  .193 & (13)\\
 V1699 Ori      &    YSO & $  -.209$ &  .428 & $  -.144$ &  .399 & $   .062$ &  .267 & (8)\\
 GMR G          &    YSO & $  -.048$ &  .139 & $   .745$ &  .188 & $  -.242$ &  .067 & (8)\\
 GMR F          &    YSO & $  -.250$ &  .128 & $   .231$ &  .164 & $  -.048$ &  .074 & (8)\\
 Parenago 1469  &    YSO & $   .026$ &  .107 & $  -.047$ &  .120 & $   .013$ &  .049 & (8)\\
 Parenago 1540  &    PMS & $   .162$ &  .128 & $   .212$ &  .109 & $  -.096$ &  .063 & (8)\\
 Parenago 1724  &    YSO & $   .199$ &  .209 & $   .153$ &  .170 & $  -.078$ &  .057 & (8)\\
 Parenago 1778  &    YSO & $   .332$ &  .499 & $   .284$ &  .728 & $  -.116$ &  .312 & (8)\\
 Parenago 1955  &    YSO & $ -2.249$ &  .694 & $ -4.038$ & 1.053 & $  -.594$ &  .215 & (8)\\
 Parenago 2148  &    YSO & $  2.276$ &  .347 & $   .909$ &  .530 & $   .606$ &  .429 & (8)\\
 V621 Ori       &    YSO & $   .475$ &  .463 & $  -.335$ &  .293 & $   .269$ &  .115 & (8)\\
 HIP 26220      & HAe/Be & $ -3.274$ &  .187 & $  2.653$ &  .184 & $  -.253$ &  .145 & (8)\\
 HIP 26314      &  B3III & $   .392$ &  .142 & $   .366$ &  .160 & $   .170$ &  .076 & (8)\\
 RW Lep         &   Mira & $  1.139$ &  .634 & $ -2.792$ &  .724 & $   .735$ &  .209 & (14)\\
 HD 294300      &  T Tau & $  7.695$ &  .682 & $ -7.858$ & 1.376 & $  -.514$ &  .373 & (8)\\
  \hline
    \end{tabular}\end{center}
    \end{table}
%%%%%%%%%%%% t2
 \begin{table}[p]
  {\small Table 1. Contd.}
  \begin{center}  \label{t:02}\small
  \begin{tabular}{|l|c|r|r|r|r|r|r|c|c|}\hline
 Star           &  Type or & $\Delta \mu_\alpha\cos\delta,$ & $\sigma_{\Delta\mu_\alpha\cos\delta},$ & $\Delta \mu_\delta,$ & $\sigma_{\Delta\mu_\delta},$ & $\Delta \pi,$ & $\sigma_{\Delta\pi},$ & Ref \\
                & spectrum &      mas yr$^{-1}$   &              mas yr$^{-1}$  &         mas yr$^{-1}$   &               mas yr$^{-1}$   &           mas &                   mas & \\\hline

 TYC 5346-538-1 &   B8.1 & $   .147$ &  .171 & $   .188$ &  .290 & $   .045$ &  .091 & (8)\\
 HD 290862      &   B3/5 & $  -.607$ &  .291 & $ -1.482$ &  .836 & $  -.020$ &  .549 & (8)\\
 U Lyn          &   Mira & $ -2.257$ &  .607 & $  -.297$ &  .602 & $  -.690$ &  .232 & (15)\\
 R UMa          &   Mira & $  1.436$ &  .551 & $   .691$ &  .517 & $   .075$ &  .208 & (16)\\
 RT Vir           &  M8III &       --- &   --- &       --- &   --- & $ -2.367$ &  .320 & (17)\\
 FV Boo           &   Mira &       --- &   --- &       --- &   --- & $  -.397$ &  .191 & (18)\\
 S Crt            &  M6III & $  -.869$ &  .327 & $   .460$ &  .268 & $   .316$ &  .195 & (19)\\
 R Cas            &   Mira & $  1.400$ & 2.384 & $  1.660$ & 1.786 & $  -.328$ & 1.965 & (20)\\
 RX Boo           & M7.5/8 & $ -3.572$ & 1.178 & $  1.809$ & 2.432 & $   .519$ &  .583 & (21)\\
 S CrB            &   Mira & $ -1.671$ &  .526 & $  1.172$ &  .467 & $  -.038$ &  .366 & (22)\\
 U Her            &   Mira & $  -.261$ &  .360 & $  -.911$ &  .392 & $ -1.991$ &  .628 & (22)\\
 WLY 2--11        &  T Tau & $  2.725$ &  .361 & $  -.388$ &  .301 & $   .231$ &  .181 & (23)\\
 YLW 24           &  T Tau & $  -.083$ &  .213 & $   .268$ &  .148 & $  -.143$ &  .166 & (23)\\
 DoAr21           &  T Tau & $  -.554$ &  .269 & $   .155$ &  .176 & $   .061$ &  .243 & (23)\\
 rho Oph S1       &  T Tau & $  -.120$ &  .254 & $  3.163$ &  .167 & $   .917$ &  .145 & (23)\\
 VSSG11           &  T Tau & $   .739$ & 1.118 & $ 14.217$ &  .776 & $  -.523$ &  .521 & (23)\\
 DROXO 71         &    PMS & $  -.799$ &  .640 & $  1.376$ &  .525 & $  -.812$ &  .312 & (23)\\
 SFAM 87          &  T Tau & $  1.143$ &  .142 & $ -2.653$ &  .111 & $   .345$ &  .115 & (23)\\
 GWAYL 5          &  T Tau & $ -1.188$ &  .568 & $   .532$ &  .428 & $  -.669$ &  .342 & (23)\\
 DoAr51           &  T Tau & $  -.396$ & 1.071 & $  1.572$ &  .726 & $   .265$ &  .387 & (23)\\
 VX Sgr           &    RSG & $  2.091$ &  .883 & $  3.691$ &  .875 & $   .147$ &  .232 & (24)\\
 $[$GFM2007$]$ 11 &    YSO & $  -.654$ &  .139 & $   .862$ &  .170 & $  -.072$ &  .109 & (25)\\
 $[$GFM2007$]$ 65 &    YSO & $  3.397$ & 1.873 & $  2.574$ & 2.086 & $  -.780$ &  .852 & (25)\\
 W 40 IRS 5       &     B1 & $   .360$ &  .404 & $  -.487$ &  .360 & $  -.249$ &  .221 & (25)\\
 W 40 IRS 1c      &    YSO & $ -3.102$ &  .892 & $  2.848$ &  .752 & $   .840$ &  .476 & (25)\\
 $[$KGF2010$]$ 133&    YSO & $ -1.177$ &  .472 & $  -.881$ &  .520 & $  -.379$ &  .245 & (25)\\
 PN K 3--35       &     PN & $   .545$ &  .157 & $  2.459$ &  .194 & $   .123$ &  .131 & (26)\\
 RR Aql           &   Mira & $  3.713$ &  .883 & $  1.077$ &  .614 & $  1.566$ &  .499 & (22)\\
 Cyg X--1         &     BH & $  -.102$ &  .077 & $   .229$ &  .132 & $  -.117$ &  .046 & (27)\\
 IRAS 20126+4104  &    YSO & $ -1.853$ &  .790 & $ -5.558$ &  .861 & $   .275$ &  .369 & (28)\\
 IRAS 20143+3634  &    YSO & $  -.123$ &  .193 & $  1.447$ &  .454 & $  -.047$ &  .080 & (29)\\
 V404 Cyg         &     BH & $  -.729$ &  .176 & $  -.205$ &  .176 & $   .021$ &  .103 & (30)\\
 HIP 101341       &  O6.5+ & $ -1.443$ &  .985 & $  3.075$ & 1.282 & $   .028$ &  .227 & (31)\\
 NML Cyg          &    RSG & $  1.282$ & 1.260 & $  3.727$ & 1.310 & $   .906$ &  .570 & (32)\\
 UX Cyg           &   Mira & $  3.381$ &  .810 & $   .254$ & 1.621 & $  -.364$ &  .178 & (33)\\
 SS Cyg          & Df Nova & $  -.047$ &  .133 & $   .209$ &  .117 & $  -.076$ &  .130 & (34)\\
 IRAS 22480+6002  &    RSG & $  -.075$ &  .354 & $  -.250$ &  .212 & $   .079$ &  .082 & (35)\\
 IM Peg           & RS CVn & $   .111$ &  .164 & $   .419$ &  .159 & $  -.320$ &  .114 & (36)\\
 R Aqr            &  M6.5e & $ -9.800$ &  .632 & $ -1.239$ &  .593 & $ -1.578$ &  .847 & (37)\\
 PZ Cas           &    RSG & $   .590$ &  .232 & $   .192$ &  .320 & $   .064$ &  .085 & (38)\\
  \hline
    \end{tabular}\end{center}
    \end{table}
%%%%%%%%%%%%%%%%%%%%%%%%%%%%%%%%%%%%%%%%%%%%% t2
%%%%%%%%%%%% t3
 \begin{table}[p]
  {\small Table 1. end.}
  \begin{center}  \label{t:03}\small
  \begin{tabular}{|l|c|r|r|r|r|r|r|c|c|}\hline
 Star           &  Type or & $\Delta \mu_\alpha\cos\delta,$ & $\sigma_{\Delta\mu_\alpha\cos\delta},$ & $\Delta \mu_\delta,$ & $\sigma_{\Delta\mu_\delta},$ & $\Delta \pi,$ & $\sigma_{\Delta\pi},$ & Ref \\
                & spectrum &      mas yr$^{-1}$   &              mas yr$^{-1}$  &         mas yr$^{-1}$   &               mas yr$^{-1}$   &           mas &                   mas & \\\hline

 UX Ari           & RS CVn & $  5.089$ &  .525 & $  2.111$ &  .411 & $   .443$ &  .452 & (39)\\
 HR 1099          & RS CVn & $ -1.304$ &  .355 & $  -.082$ &  .332 & $  -.127$ &  .478 & (39)\\
 HIP 79607        & RS CVn & $ -1.275$ &  .104 & $  -.265$ &  .154 & $   .205$ &  .119 & (39)\\
 HD 199178        &  G5III & $  -.277$ &  .415 & $   .498$ &  .435 & $   .312$ &  .332 & (39)\\
 AR Lac           & RS CVn & $  -.110$ &  .137 & $   .160$ &  .195 & $  -.537$ &  .371 & (39)\\
 AM Her           &  polar & $   .063$ &  .223 & $  -.784$ &  .183 & $   .105$ &  .082 & (40)\\
 W Hya            &   Mira & $ -7.533$ & 2.418 & $ -4.408$ & 3.234 & $ -4.089$ & 2.497 & (20)\\
 VY CMa           &    RSG & $  3.726$ & 1.865 & $ -9.074$ & 1.847 & $ -6.772$ &  .827 & (41)\\
  \hline
    \end{tabular}
    \end{center}
 {\small
Mira—Mira Ceti variable; RSG—red supergiant; RS CVn—Canes Venatici
variable; BY Dra—BY Draconis variable; T Tau— T Tauri variable;
PMS—pre-main-sequence star; HAe/Be—Herbig Ae/Be star; YSO—young
stellar object; PN—planetary nebula; Df Nova—dwarf nova; BH—one of
the binary components is a black hole; WR—Wolf–Rayet star.

(1) Nyu et al. (2011); (2) Asaki et al. (2010); (3) Melis et al.
(2014); (4) Torres et al. (2012); (5) Galli et al. (2018); (6)
Loinard et al. (2007); (7) Dhawan et al. (2006); (8) Kounkel et
al. (2017); (9) Rivera et al. (2015); (10) Abbuhl et al. (2015);
(11) Dzib and Rodriguez (2009); (12) Deller et al. (2008); (13)
Nakagawa et al. (2014); (14) Kamezaki et al. (2014); (15) Kamezaki
et al. (2016a); (16) Nakagawa et al. (2016); (17) Zhang et al.
(2017); (18) Kamezaki et al. (2016b); (19) Nakagawa et al. (2008);
(20) Vlemmings et al. (2003); (21) Kamezaki et al. (2012); (22)
Vlemmings, and van Langevelde (2007); (23) Ortiz-Leon et al.
(2017a); (24) Xu et al. (2018); (25) Ortiz-Leon et al. (2017b);
(26) Tafoya et al. (2011); (27) Reid et al. (2011); (28) Xu et al.
(2013); (29) Burns et al. (2014); (30) Miller-Jones et al. (2009);
(31) Dzib et al. (2013); (32) Zhang et al. (2012a); (33) Kurayama
et al. (2005); (34) Miller-Jones et al. (2013); (35) Imai et al.
(2012); (36) Ratner et al. (2012); (37) Min et al. (2014); (38)
Kusuno et al. (2013); (39) Lestrade et al. (1999); (40)
Gawro\'nski et al. (2018); (41) Zhang et al. (2012b).
  }
    \end{table}
%%%%%%%%%%%%%%%%%%%%%%%%%%%%%%%%%%%%%%%%%%%%% t3

 \subsection*{DATA}
In this paper we collected the VLBI observations of stellar
trigonometric parallaxes and proper motions performed and
published by various research teams. For example, these include
the Japanese VERA (VLBI Exploration of Radio Astrometry) project
devoted to the observations of H$_2$O masers at 22.2 GHz (Hirota
et al. 2008) and a number of SiO masers at 43 GHz (Kim et al.
2008). Methanol (CH$_3$OH) and H$_2$O masers are observed in the
USA on the VLBA (Reid et al. 2009). Similar observations are also
performed within the framework of the European VLBI network (Rygl
et al. 2010). The VLBI observations of radio stars in continuum at
8.4 GHz are also carried out with the same goals (Torres et al.
2012).

Table 1 gives the proper motion and trigonometric parallax
differences for 88 stars. The stars have a different evolutionary
status. Some of them are very young stars with maser emission
(H$_2$O and CH$_3$OH masers). Asymptotic giant branch stars
observed as OH, H$_2$O, and SiO masers constitute the other part
of the sample. Quite a few sources were observed in continuum.
This is true for such objects as pulsars, Wolf–Rayet stars,
systems with black holes, and a number of T Tauri stars.

In our previous paper (Bobylev 2010) we used 23 radio stars from
this list to study the tie-in of the Hipparcos catalogue (1997) to
the inertial reference frame. In this paper the list was expanded
significantly both through an increase in the number of VLBI
observations and owing to a great density of the Gaia DR2
catalogue.

The first column in the table gives the names of the radio stars
using which they are easily found in the SIMBAD electronic search
system. The second column lists the types of the stars or their
spectral types. The next columns provide the proper motion and
trigonometric parallax differences. The dispersions of the
differences are given for each type of differences. For example,
for the proper motions the formula to calculate the dispersions of
the differences is as follows:
 \begin{equation}
  \sigma_{\Delta_\mu} = \sqrt{\sigma^2_{\mu_{Gaia}}+\sigma^2_{\mu_{VLBI}} },
 \label{differences}
 \end{equation}
for the dispersions of the parallax differences the expression is
similar in form after an appropriate substitution.

There are no parallax differences for eight stars in Table~1. This
is either due to negative parallaxes in the Gaia DR2 catalogue or
the absence (for example, for Wolf–Rayet stars) of VLBI
measurements. At the same time, we used these eight stars to
analyze the proper motion differences. For two stars, RT Vir and
FV Boo, there are data only on their VLBI parallax measurements.

It can be seen from Table 1 that several stars have differences
that differ significantly from the expected zero. For example,
these include the stars R Aqr with $\Delta\mu_\alpha
\cos\delta=-9.800\pm0.632$~mas yr$^{-1},$ VSSG11 with
$\Delta\mu_\delta=14.217\pm0.776$~mas yr$^{-1},$ or VY CMa with
$\Delta\pi=-6.772\pm0.827$~mas. Note that the presence of a long
tail in the distribution of radio source position differences was
established by Petrov and Kovalev (2017) when analyzing a large
sample of quasars from the Gaia catalogue with VLBI measurements.

When the optical and radio images of stars are compared, the size
and pattern of the radio-emitting region can play an important
role. The supergiant S Per can serve as an example of a ``good'',
symmetric radio image. As can be seen from Fig. 5 in Asaki et al.
(2010), more than 40maser spots are distributed quite uniformly in
a region with a radius of about 50 mas, while, according to Fig.
10 in the cited paper, the residual velocity vectors excellently
pinpoint the position of the image center. As can be seen from our
table, all differences for the star S Per are close to zero.

On the other hand, the radio emission can be associated with the
jets or vast disk structures surrounding the radio star. In that
case, the probability of the appearance of a significant offset
when comparing the optical and radio images of a star is great.

Finally, the optical image of a radio star can also be asymmetric.
The well-known star VY CMa can serve as such an example. This is a
red supergiant; the star has a record size. It is actually a
presupernova and is surrounded by a nebula with a highly
asymmetric shape.

All of this necessitates using constraints on the differences
being investigated when solving our problems. Such constraints
were selected through several iterations to eliminate the largest
discrepancies.

%%%%%%%%%%%%%%%%%%%%%%%% FIG.1:
 \begin{figure}[t]{\begin{center}
 \includegraphics[width=0.5\textwidth]{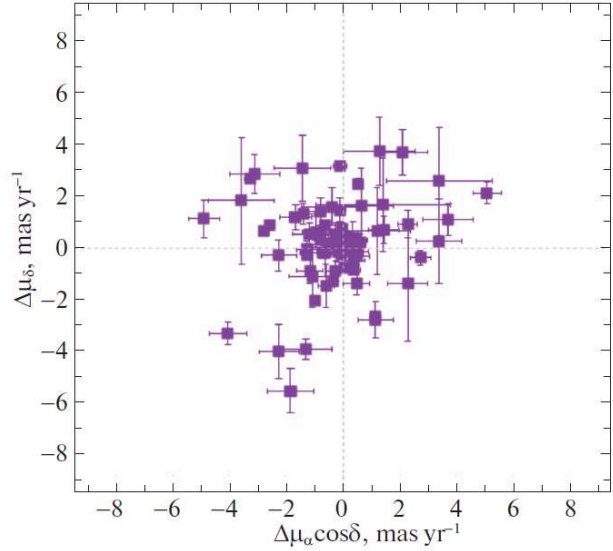}
 \caption{Gaia--VLBI stellar proper motion differences.} \label{f-differ}
\end{center}}\end{figure}
%%%%%%%%%%%%%%%%%%%%%%%%%%%%%%%%%%%%%%%%%%%%%%%%%%%%%%%%%%%%%
%%%%%%%%%%%%%%%%%%%%%%%% FIG.2:
\begin{figure}[t]{\begin{center}
   \includegraphics[width=0.95\textwidth]{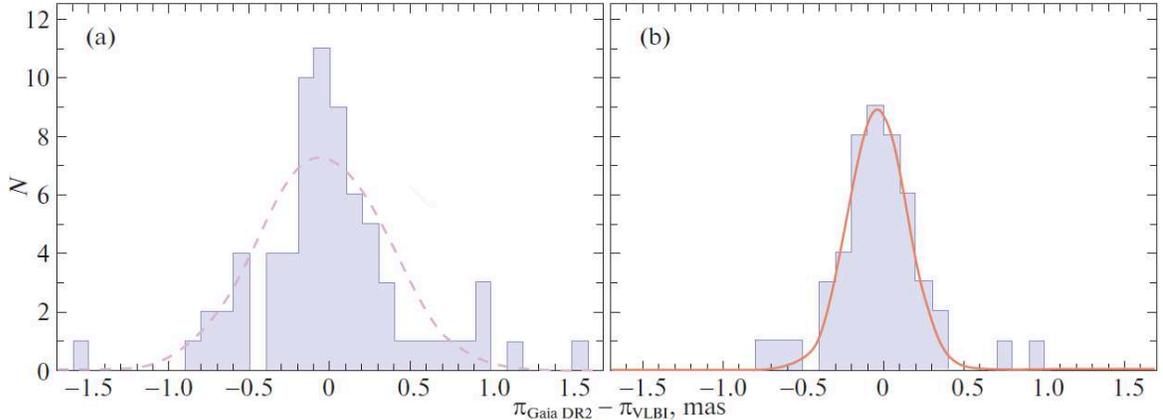}
\caption{The histogram of Gaia--VLBI parallax differences
constructed from all differences: a Gaussian with an expectation
value of $-0.30$~mas and a dispersion of 0.40 mas (a) and a
Gaussian with the constraint on the difference
$\sigma_{\Delta\pi}<0.25$~mas (here it has an expectation value of
$-0.35$~mas and a dispersion of 0.18 mas) (b) are
shown.}\label{f-hist}
\end{center}}\end{figure}
%%%%%%%%%%%%%%%%%%%%%%%%%%%%%%%%%%%%%%%%%%%%%%%%%%%%%%%%%%%%%
%%%%%%%%%%%%%%%%%%%%%%%% FIG.3:
\begin{figure}[t]{\begin{center}
   \includegraphics[width=0.5\textwidth]{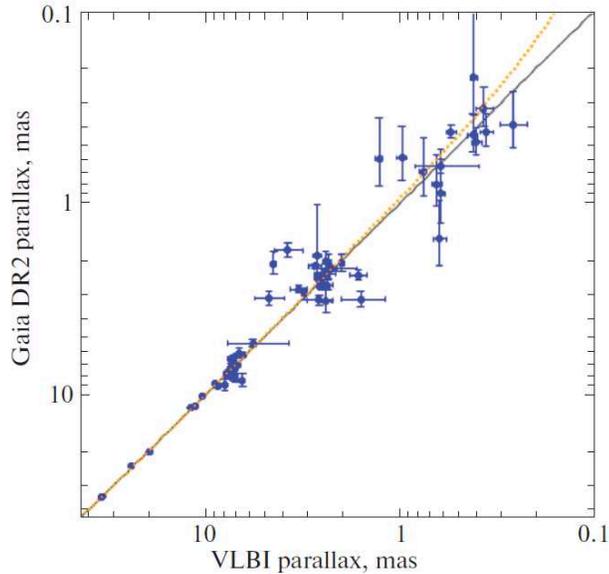}
\caption{ Parallaxes of the radio stars from the Gaia DR2
catalogue versus their parallaxes measured by VLBI; the solid and
dotted lines correspond to a correlation with a coefficient of 1
and the solution (8), respectively.} \label{f-par}
\end{center}}\end{figure}
%%%%%%%%%%%%%%%%%%%%%%%%%%%%%%%%%%%%%%%%%%%%%%%%%%%%%%%%%%%%%

 \section*{RESULTS}\label{results}
 \subsection*{Comparison of the Proper Motions}\label{mu}
We use the following coupling equations to determine the three
angular velocities of mutual rotation of the two frames around the
equatorial coordinate axes $\omega_x,\omega_y,\omega_z$:
 \begin{equation}
 \begin{array}{lll}
 \Delta\mu_\alpha\cos\delta=-\omega_x\cos\alpha\sin\delta -\omega_y\sin\alpha\sin\delta +\omega_z\cos\delta, \\
 \Delta\mu_\delta=\omega_x\sin\alpha -\omega_y\cos\alpha,
 \label{DR2-VLBI}
 \end{array}
 \end{equation}
where the Gaia--VLBI differences are on the left-hand sides of the
equations. We use the stellar proper motion differences whose
absolute values do not exceed 6 mas yr$^{-1}.$ There are a total
of 81 such differences; their distribution is given in Fig.~1.

As can be seen from the table, the data are unequally accurate.
Therefore, we solve the system of conditional equations (2) both
with unit weights $(p=1)$ and with weights inversely proportional
to the measurement errors
 \begin{equation}
 p = 1/\sqrt{\sigma^2_{\mu_{Gaia}}+\sigma^2_{\mu_{VLBI}} },
 \label{p-not1}
 \end{equation}
where the dispersions $\sigma_\Delta$ listed in the corresponding
columns of the table are in the denominator (see Eq. (1)).

Having solved the system of 162 conditional equations (2) by the
least-squares method with unit weights, we obtained the rotation
components
 \begin{equation}
 \begin{array}{lll}
 \omega_x=-0.44\pm0.20~\hbox{mas yr$^{-1}$}, \\
 \omega_y=-0.05\pm0.29~\hbox{mas yr$^{-1}$}, \\
 \omega_z=-0.27\pm0.21~\hbox{mas yr$^{-1}$}.
 \label{w1w2w3-p1}
 \end{array}
 \end{equation}
At the same time, with weights (3) we obtained the rotation
components
 \begin{equation}
 \begin{array}{lll}
 \omega_x=-0.14\pm0.15~\hbox{mas yr$^{-1}$}, \\
 \omega_y=+0.03\pm0.22~\hbox{mas yr$^{-1}$}, \\
 \omega_z=-0.33\pm0.16~\hbox{mas yr$^{-1}$},
 \label{w1w2w3-p2}
 \end{array}
 \end{equation}
where $\omega_x$ decreased greatly compared to the solution (4);
the errors in the parameters being determined also decreased.

 \subsection*{Comparison of the Parallaxes}\label{PAR}
To compare the parallaxes, we use 75 stars selected in such a way
that the relative parallax errors from the Gaia DR2 catalogue and
the VLBI parallax errors do not exceed 50\%.

First, we found the mean $\Delta\pi=-0.030\pm0.073~(0.404)$~mas
from the Gaia-VLBI parallax differences. The mean was calculated
with unit weights, the error of the mean calculated from the
formula $\sqrt{\sum{(x-{\overline x})^2}/n(n-1)},$ is given, and
the dispersion $\sigma=\sum{(x-{\overline x})^2}/n$ (here the
square of the rms deviation) is given in parentheses. Then, we
calculated the weighted mean with weights (3)
 \begin{equation}
 \Delta\pi=-0.038\pm0.046~(0.156)~\hbox{mas},
 \label{weighted}
 \end{equation}
where the error of the weighted mean is given and the
corresponding dispersion is given in parentheses. We see that the
errors and dispersions differ greatly. This effect is explained by
the fact that we used significantly inhomogeneous data. Very broad
distribution wings can be seen already from the distribution of
stellar proper motion differences (Fig.~1), namely (a) a central
clump that can be described by a Gaussian with a small dispersion
and (b) broad wings that can be described by a Gaussian with a
considerably larger dispersion.

The effect is more pronounced in the distribution of stellar
parallax differences. The histogram of differences for 75 stars is
presented in Fig. 2a. This figure shows a Gaussian with an
expectation value of $-0.30$~mas and a dispersion of $0.40$~mas
that poorly describes the distribution. Two Gaussians with
significantly differing dispersions would be better suited for the
description of this distribution. However, we did otherwise. To
construct the histogram in Fig. 2b, we used 49 stars that were
selected under constraints on the error in the differences (see
(1) and the table): $\sigma_{\Delta\pi}<0.25$~mas. The parameters
of the Gaussian found (an expectation value of $-0.35$~mas and a
dispersion of 0.18~mas) are now in excellent agreement with the
result (6). On this basis we conclude that the application of
weights (3) gives a result consistent with the available data;
this approach allows the entire set of available data to be used.

To determine the scale factor $b,$ we set up a system of
conditional linear equations
 \begin{equation}
 \begin{array}{lll}
  \pi_{Gaia}=a+b\cdot\pi_{VLBI},
 \label{EQ-1}
 \end{array}
\end{equation}
from the solution of which we can estimate two parameters, a and
b. As above, we use 75 stars with relative parallax errors less
than 50\%. Solving the system of conditional equations (7) by the
least-squares method with weights (3) yields the following result:
 \begin{equation}
 \begin{array}{lll}
  a=-0.048\pm0.059~\hbox {mas},\\
  b=+1.002\pm0.007.
 \label{linear}
 \end{array}
\end{equation}
In Fig. 3 the parallaxes of the radio stars from the Gaia DR2
catalogue are plotted against their VLBI parallaxes. The scales
are clearly seen to be virtually identical within about 3~kpc of
the Sun, and only at greater distances does the Gaia DR2 parallax
scale become longer than the VLBI parallax one.

 \subsection*{DISCUSSION}
Liu et al. (2017) studied the frame of the Gaia DR1 catalogue
(Brown et al. 2016). In particular, the TGAS (Tycho-Gaia
Astrometric Solution) version was compared with the Tycho2
catalogue (H$\o$g et al. 2000) and the version of the Hipparcos
catalogue (1997) improved by van Leeuwen (2007) using the model of
solid-body rotation (2). These authors found the rotation vector
components
$(\omega_x,\omega_y,\omega_z)=(0.008,0.010,-0.014)\pm(0.007,0.007,0.009)$
mas yr$^{-1}$ from the Hipparcos-TGAS proper motion differences
for $\sim$87 000 stars and
$(\omega_x,\omega_y,\omega_z)=(0.011,0.013,0.024)\pm(0.004,0.004,0.005)$
mas yr$^{-1}$ from the Tycho2-TGAS proper motion differences for
$\sim$2 million stars. Thus, Liu et al. (2017) revealed no
significant mutual rotations between these frames.

However, based on the Ogorodnikov-Milne model, Liu et al. (2017)
performed a kinematic analysis of $\sim$23 000 K--M giants from
the TGAS catalogue and found nonzero components pointing to a
possible residual rotation in the Gaia DR1 frame or the presence
of problems in the kinematic model. The rotation components were
found to be $\omega_{Y_G}=-0.38\pm0.15$ mas yr$^{-1}$ and
$\omega'_{Y_G}=-0.29\pm0.19$ mas yr$^{-1}$, which are interpreted
as an additional rotation around the Galactic $Y$~axis.

Note the paper by Fedorov et al. (2017), where it was found from a
comparison of the stellar proper motions from the Gaia DR1
catalogue with a number of ground-based catalogues based on the
model (2) that the component $\omega_y$ changes dramatically from
$+0.5$ to $-1.5$~mas yr$^{-1}$ with magnitude. In our case (5)
this component is small, $\omega_y=0.03\pm0.22$ mas yr$^{-1}$.

It has been shown by Lindegren et al. (2018) that the optical
reference frame defined by Gaia DR2 is aligned with ICRS and is
non-rotating with respect to the quasars to within 0.15 mas
yr$^{-1}$. Since a large number of stars were used, the random
errors of rotational parameters are small, less than 10\%. The
dependence of $\omega_x,\omega_y,\omega_z$ on magnitude is clearly
seen from Fig. 4 of cited publication. For example, for
$G\approx10^m$, which is typical for the sample of stars
considered in this paper, we will have
$(\omega_x,\omega_y,\omega_z)\approx(0.1,-0.1,-0.15)$ mas
yr$^{-1}$. We see good agreement of these values with our
estimates (5).

As has already been noted in the Introduction, from a comparison
with the Gaia DR2 data for 89 detached eclipsing binaries Stassun
and Torres (2018) found a correction
$\Delta\pi=-0.082\pm0.033$~mas. Here the dispersion of the
Gaussian 0.033~mas should be compared with our value of 0.156 mas
in the solution (6). These stars are interesting in that they were
selected from published data using very rigorous criteria imposed
on the photometric characteristics. As a result, the relative
errors in the stellar radii, effective temperatures, and
bolometric luminosities (from which the distances are estimated)
do not exceed 3\%. The spectral types of the stars in this sample
lie in a wide range, from late O to M; most of the stars belong to
the main sequence and there are also a few giants. According to
Stassun and Torres (2016), the relative parallax errors for
eclipsing binaries, on average, do not exceed 5\% and do not
depend on the distance.

Riess et al. (2018) estimated $\Delta\pi=-0.046\pm0.013$~mas based
on a sample of 50 long-period Cepheids by comparing their
parallaxes with those from the Gaia DR2 catalogue. They used the
photometric characteristics of these Cepheids measured onboard the
Hubble Space Telescope. Interestingly, relative to the highly
accurate calibration scale of Riess et al. (2016), in which the
relative Cepheid distance errors are 1--2\%, these authors
determined the scale factor $b=1.006\pm0.033$ that differs little
from that found by us in the solution (8).

One might expect that the stellar parallaxes from the Gaia DR1 and
DR2 catalogues do not greatly differ systematically. For example,
based on a kinematic analysis of stars from the Gaia TGAS
catalogue, Bobylev and Bajkova (2018) concluded that the distances
to them calculated from their trigonometric parallaxes do not
require using any additional correction factor. This conclusion is
also confirmed by our study with regard to the stellar parallaxes
from the Gaia DR2 catalogue.

Zinn et al. (2018) found $\Delta\pi=-0.083\pm0.002$~mas by
comparing the distances of $\sim $3000 giants from the APOKAS-2
catalogue (Pinsonneault et al. 2018) with the Gaia DR2 data. The
distances to these stars belonging to the red giant clump were
calculated from asteroseismology. According to these authors, here
the parallax errors are approximately equal to the estimation
errors of the stellar radius and are, on average, 1.5\%. Such
small errors in combination with a huge number of stars allowed
$\Delta\pi$ to be determined with a high accuracy.

Young stars from the Gould Belt, the distances to which have been
measured by VLBI, constitute a significant fraction of our sample.
Using data on 55 such stars (they are all presented in our table
as PMS, YSO, and T Tau), Kounkel et al. (2018) found the following
parameters based on relation (7): $a=-0.073\pm0.034$~mas and
$b=+0.9947\pm0.0066.$ The value of these parameters are in
excellent agreement with our estimates (8).

 \subsection*{CONCLUSIONS}
Based on published data, we produced a sample of 88 radio stars
for which there are both trigonometric parallax measurements in
the Gaia DR2 catalogue and VLBI measurements.

A new estimate of the systematic offset between the optical and
radio frames of the parallaxes,
$\Delta\pi=-0.038\pm0.046~(0.156)$~mas, was obtained by analyzing
the Gaia--VLBI trigonometric parallax differences for the radio
stars. If the VLBI parallaxes are assumed to be more accurate,
then the correction found should be added to the parallaxes from
the Gaia DR2 catalogue. In this case, the distances to the stars
calculated from the corrected Gaia DR2 parallaxes slightly
decrease, i.e., the stars will become closer to the Sun.

The scale factor $b,$ whose value differs from 1 by no more than
1\%, is determined with confidence. Such a situation is observed
within 3 kpc of the Sun, and only at greater distances is the Gaia
DR2 parallax scale slightly extended compared to the VLBI parallax
one.

Based on the model of solid-body mutual rotation, we determined
the rotation vector components in equatorial coordinates from the
Gaia-VLBI proper motion differences for radio stars,
$(\omega_x,\omega_y,\omega_z)=(-0.14,0.03,-0.33)\pm(0.15,0.22,0.16)$
mas yr$^{-1}.$

 \subsubsection*{ACKNOWLEDGMENTS}
I am grateful to the referee for the useful remarks that
contributed to an improvement of the paper. This work was
supported by Basic Research Program P--28 of the Presidium of the
Russian Academy of Sciences, the subprogram ``Cosmos: Studies of
Fundamental Processes and their Interrelations''.

 \bigskip
 \bigskip\medskip{\bf REFERENCES}
{\small

1. E. Abbuhl, R. L. Mutel, C. Lynch, and M. Guedel, Astrophys. J.
811, 33 (2015).

2. Y. Asaki, S. Deguchi, H. Imai, K. Hachisuka, M. Miyoshi, and M.
Honma, Astrophys. J. 721, 267 (2010).

3. V. V. Bobylev, Astron. Lett. 41, 156 (2015).

4. V. V. Bobylev and A. T. Bajkova, Astron. Lett. 44, 184 (2018).

5. A. G. A. Brown, A. Vallenari, T. Prusti, J. de Bruijne, F.
Mignard, R. Drimmel, C. Babusiaux, C. A. L. Bailer-Jones, et al.
(Gaia Collab.), Astron. Astrophys. 595, A2 (2016).

6. A. G. A. Brown, A. Vallenari, T. Prusti, J. de Bruijne, C.
Babusiaux, C. A. L. Bailer-Jones, M. Biermann, D. W. Evans, et al.
(Gaia Collab.), Astron. Astrophys. 616, 1 (2018).

7. R. A. Burns, Y. Yamaguchi, T. Handa, T. Omodaka, T. Nagayama,
A. Nakagawa, M. Hayashi, T. Kamezaki, et al., Publ. Astron. Soc.
Jpn. 66, 102 (2014).

8. S. Casertano, A. G. Riess, B. Bucciarelli, and M. G. Lattanzi,
Astron. Astrophys. 599, 67 (2017).

9. A. T. Deller, J. P. W. Verbiest, S. J. Tingay, and M. Bailes,
Astrophys. J. 685, L67 (2008).

10. V. Dhawan, A. Mioduszewski, and M. Rupen, in {\it Proceedings
of the 6th Microquasar Workshop: Microquasars and Beyond,
September 18–22, 2006, Como, Italy,} (2006), p.52.1.

11. S. A. Dzib and L. F. Rodriguez, Rev. Mex. Astron. Astrofis.
45, 3 (2009).

12. S. A. Dzib, L. F. Rodriguez, L. Loinard, A. J. Mioduszewski,
G. N. Ortiz-Le\'on, and A. T. Araudo, Astrophys. J. 763, 139
(2013).

13. S. Dzib, L. Loinard, L. F. Rodriguez, A. J. Mioduszewski, G.
N. Ortiz-Le\'on, M. A. Kounkel, G. Pech, J. L. Rivera, et al.,
Astrophys. J. 801, 91 (2015).

14. P. N. Fedorov, V. S. Akhmetov, and A. B. Velichko, Mon. Not.
R. Astron. Soc. 476, 2743 (2017).

15. P. A. B. Galli, L. Loinard, G. N. Ortiz-Le\'on, M. Kounkel, S.
A. Dzib, A. J. Mioduszewski, L. F. Rodriguez, L. Hartmann, et al.,
Astrophys. J. 859, 33 (2018).

16. M. P. Gawro\'nski, K. Go\'zdziewski, K. Katarzy\'nski, and G.
Rycyk, Mon. Not. R. Astron. Soc. 475, 1399 (2018).

17. The HIPPARCOS and Tycho Catalogues, ESA SP--1200 (1997).

18. T. Hirota, T. Bushimata, Y. K. Choi, M. Honma, H. Imai, I.
Hiroshi, K. Iwadate, T. Jike, et al., Publ. Astron. Soc. Jpn. 60,
37 (2008).

19. E. H$\o$g C. Fabricius, V. V. Makarov, S. Urban, T. Corbin, G.
Wycoff, U. Bastian, P. Schwekendiek, and A. Wicenec, Astron.
Astrophys. 355, L27 (2000).

20. D. Huber, J. Zinn, M. Bojsen-Hansen, M. Pinsonneault, C.
Sahlholdt, A. Serenelli, V. S. Aguirre, K. Stassun, et al.,
Astrophys. J. 844, 102 (2017).

21. H. Imai, N. Sakai, H. Nakanishi, H. Sakanoue, M. Honma, and T.
Miyaji, Publ. Astron. Soc. Jpn. 64, 142 (2012).

22. W.-C. Jao, T. J. Henry, A. R. Riedel, J. G.Winters, K. J.
Slatten, and D. R. Gies, Astrophys. J. Lett. 832, L18 (2016).

23. T. Kamezaki, A. Nakagawa, T. Omodaka, T. Kurayama, H. Imai, D.
Tafoya, M. Matsui, and Y. Nishida, Publ. Astron. Soc. Jpn. 64, 7
(2012).

24. T. Kamezaki, T. Kurayama, A. Nakagawa, T. Handa, T. Omodaka,
T. Nagayama, H. Kobayashi, and M. Shizugami, Publ. Astron. Soc.
Jpn. 66, 107 (2014).

25. T. Kamezaki, A. Nakagawa, T. Omodaka, T. Handa, K.-I. Inoue,
T. Kurayama, H. Kobayashi, T. Nagayama, et al., Publ. Astron. Soc.
Jpn. 68, 71 (2016a).

26. T. Kamezaki, A. Nakagawa, T. Omodaka, K.-I. Inoue, J. O.
Chibueze, T. Nagayama, Y. Ueno, and N. Matsunaga, Publ. Astron.
Soc. Jpn. 68, 75 (2016b).

27. M. K. Kim, T. Hirota, M. Honma, H. Kobayashi, T. Bushimata, Y.
K. Choi, H. Imai, K. Iwadate, et al., Publ. Astron. Soc. Jpn. 60,
991 (2008).

28. M. Kounkel, L. Hartmann, L. Loinard, G. N. Ortiz-Le\'on, A. J.
Mioduszewski, L. F. Rodriguez, R. M. Torres, G. Pech, et al.,
Astrophys. J. 834, 142 (2017).

29. M. Kounkel, K. Covey, G. Suarez, C. Rom\'an-Zuniga, J.
Hernandez, K. Stassun, K. O. Jaehnig, E. Feigelson, et al.,
Astron. J. 156, 84 (2018).

30. T. Kurayama, T. Sasao, and H. Kobayashi, Astrophys. J. 627,
L49 (2005).

31. K. Kusuno, Y. Asaki, H. Imai, and T. Oyama, Astrophys. J. 774,
107 (2013).

32. F. van Leeuwen, Astron. Astrophys. 474, 653 (2007).

33. J.-F. Lestrade, R. A. Preston, D. L. Jones, R. B. Phillips, A.
E. E. Rogers, M. A. Titus, M. J. Rioja, and D. C. Gabuzda, Astron.
Astrophys. 344, 1014 (1999).

34. L. Lindegren, J. Hernandez, A. Bombrun, S. Klioner, U.
Bastian, M. Ramos-Lerate, A. de Torres, H. Steidelmuller, et al.
(Gaia Collab.), Astron. Astrophys. 616, 2 (2018).

35. N. Liu, Z. Zhu, J.-C. Liu, and C.-Y. Ding, Astron. Astrophys.
599, 140 (2017).

36. L. Loinard, R. M. Torres, A. J. Mioduszewski, L. F. Rodriguez,
R. A. Gonzalez-Lopezlira, R. Lachaume, V. Vazquez, and E.
Gonzalez, Astrophys. J. 671, 546 (2007).

37. C. Melis, M. J. Reid, A. J. Mioduszewski, J. R. Stauffer, and
G. C. Bower, Science (Washington, DC, U. S.) 345, 1029 (2014).

38. F. Mignard, S. A. Klioner, L. Lindegren, J. Hern\'andez, U.
Bastian, A. Bombrun, D. Hobbs, U. Lammers, et al. (Gaia Collab.),
Astron. Astrophys. 616, 14 (2018).

39. J. C. A. Miller-Jones, P. G. Jonker, V. Dhawan, W. Brisken,
M. P. Rupen, G. Nelemans, and E. Gallo, Astrophys. J. 706, 230
(2009).

40. J. C. A. Miller-Jones, G. R. Sivakoff, C. Knigge, E. G.
Kording, M. Templeton, and E. O. Waagen, Science (Washington, DC,
U. S.) 340, 950 (2013).

41. C. Min, N. Matsumoto, M. K. Kim, T. Hirota, K. M. Shibata,
S.-H. Cho, M. Shizugami, and M. Honma, Publ. Astron. Soc. Jpn. 66,
38 (2014).

42. A. Nakagawa, M. Tsushima, K. Ando, T. Bushimata, Y. K. Choi,
T. Hirota, M. Honma, H. Imai, et al., Publ. Astron. Soc. Jpn. 60,
1013 (2008).

43. A. Nakagawa, T. Omodaka, T. Handa, M. Honma, N. Kawaguchi, H.
Kobayashi, T. Oyama, K. Sato, et al., Publ. Astron. Soc. Jpn. 66,
101 (2014).

44. A. Nakagawa, T. Kurayama, M. Matsui, T. Omodaka, M. Honma, K.
M. Shibata, K. Sato, end T. Jike, Publ. Astron. Soc. Jpn. 66, 101
(2016).

45. D. Nyu, A. Nakagawa, M. Matsui, H. Imai, Y. Sofue, T. Omodaka,
T. Kurayama, R. Kamohara, et al., Publ. Astron. Soc. Jpn. 63, 53
(2011).

46. G. N. Ortiz-Le\'on, L. Loinard, M. A. Kounkel, S. A. Dzib, A.
J. Mioduszewski, L. F. Rodriguez, R. M. Torres, R. A.
Gonz\'alez-L\'opezlira, et al., Astrophys. J. 834, 141 (2017a).

47. G. N. Ortiz-Le\'on, S. A. Dzib, M. A. Kounkel, L. Loinard, A.
J. Mioduszewski, L. F. Rodriguez, R. M. Torres, G. Pech, et al.,
Astrophys. J. 834, 143 (2017b).

48. L.Petrov and Y. Y. Kovalev, Mon. Not. R. Astron. Soc. 467, 71
(2017).

49. M. H. Pinsonneault, Y. P. Elsworth, J. Tayar, A. Serenelli, D.
Stello, J. Zinn, S. Mathur, R. Garcia, et al., arXiv: 1804.09983
(2018).

50. N. Pradel, P. Charlot, and J.-F. Lestrade, Astron. Astrophys.
452, 1099 (2006).

51. T. Prusti, J. H. J. de Bruijne, A. G. A. Brown, A. Vallenari,
C. Babusiaux, C. A. L. Bailer-Jones, U. Bastian, M. Biermann, et
al. (Gaia Collab.), Astron. Astrophys. 595, A1 (2016).

52. M. I. Ratner, N. Bartel, M. F. Bietenholz, D. E. Lebach, J.-F.
Lestrade, R. R. Ransom, and I. I. Shapiro, Astrophys. J. Suppl.
Ser. 201, 5 (2012).

53. M. J. Reid, K. M. Menten, X. W. Zheng, A. Brunthaler, L.
Moscadelli, Y. Xu, B. Zhang, M. Sato, et al., Astrophys. J. 700,
137 (2009).

54. M. J. Reid, J. E. McClintock, R. Narayan, L. Gou, R. A.
Remillard, and J. A. Orosz, Astrophys. J. 742, 83 (2011).

55. A. G. Riess, L. Macri, S. L. Hoffmann, D. Scolnic, S.
Casertano, A. V. Filippenko, B. E. Tucker, M. J. Reid, et al.,
Astrophys. J. 826, 56 (2016).

56. A. G. Riess, S. Casertano, W. Yuan, L. Macri, B. Bucciarelli,
M. G. Lattanzi, J. W. MacKenty, J. B. Bowers, et al., Astrophys.
J. 861, 126 (2018).

57. J. L. Rivera, L. Loinard, S. A. Dzib, G. N. Ortiz-Le\'on, L.
F. Rodriguez, and R. M. Torres, Astrophys. J. 807, 119 (2015).

58. K. L. J. Rygl, A. Brunthaler, M. J. Reid, K. M. Menten, H. J.
van Langevelde, and Y. Xu, Astron. Astrophys. 511, 2 (2010).

59. K. G. Stassun and G. Torres, Astrophys. J. Lett. 831, L6
(2016).

60. K. G. Stassun and G. Torres, Astrophys. J. 862, 61 (2018).

61. D. Tafoya, H. Imai, Y. Gomez, J. M. Torrelles, N. A. Patel, G.
Anglada, L. F. Miranda, M. Honma, et al., Publ. Astron. Soc. Jpn.
63, 71 (2011).

62. R. M. Torres, L. Loinard, A. J. Mioduszewski, A. F. Boden, R.
Franco-Hernandez, W. H. T. Vlemmings, and L. F. Rodriguez,
Astrophys. J. 747, 18 (2012).

63. W. H. T. Vlemmings, H. J. van Langevelde, P. J. Diamond, H. J.
Habing, and R. T. Schilizzi, Astron. Astrophys. 407, 213 (2003).

64. W. H. T. Vlemmings and H. J. van Langevelde, Astron.
Astrophys. 472, 547 (2007).

65. S. Xu, B. Zhang, M. J. Reid, K. M. Menten, X. Zheng, and G.
Wang, Astrophys. J. 859, 14 (2018).

66. B. Zhang, M. J. Reid, K. M. Menten, X. W. Zheng, and A.
Brunthaler, Astron. Astrophys. 544, 42 (2012a).

67. B. Zhang, M. J. Reid, K. M. Menten, and X. W. Zheng,
Astrophys. J. 744, 23 (2012b).

68. B. Zhang, X. Zheng, M. J. Reid, M. Honma, K. M. Menten, A.
Brunthaler, and J. Kim, Astrophys. J. 849, 99 (2017).

69. J. C. Zinn, M. H. Pinsonneault, D. Huber, and D. Stello,
arXiv: 1805.02650 (2018).

 }

  \end{document}